\documentclass[submitted]{IEEEtran}
\usepackage{cite}
\usepackage{amsmath}
\usepackage{graphicx}
\usepackage{amssymb}
\usepackage{upgreek}
\usepackage{multirow}
\usepackage{multicol}
\usepackage{tabularx}
\usepackage[colorinlistoftodos]{todonotes}
\usepackage[breaklinks=true,colorlinks=true,linkcolor=blue,urlcolor=blue,citecolor=blue]{hyperref}
\usepackage[colorinlistoftodos]{todonotes}

\begin{document}

\title{An Evaluation of External Magnetic Flux Error in Magnet-Moving Kibble balances}
        \author{Yongchao Ma, Wei Zhao, Songling Huang, {\it Senior Member, IEEE}, Shisong Li$^\dagger$, {\it Senior Member, IEEE}
        \thanks{Yongchao Ma, Songling Huang, and Shisong Li are with the Department of Electrical Engineering, Tsinghua University, Beijing 100084, China. Wei Zhao is with the Department of Electrical Engineering, Tsinghua University, Beijing 100084, China, and the Yangtze Delta Region Institute of Tsinghua University, Jiaxing, Zhejiang 314006, China. }
        \thanks{$^\dagger$Email: shisongli@tsinghua.edu.cn.}
} 

\markboth{}{}
\maketitle

\begin{abstract}
The magnet-moving measurement scheme in Kibble balances avoids displacing force-sensitive components, such as the weighing cell, and enables a broader magnetic profile measurement range during the velocity phase. However, this mechanism introduces the risk of asymmetry in the $Bl$ measurement due to external magnetic flux, leading to a potential systematic error in the final measurement results. Using the Tsinghua tabletop Kibble balance magnet as a case study, this paper investigates the error mechanism through finite element analysis and experimental investigations. An evaluation method combining external weak-field measurements with attenuation factor analysis is proposed to assess external magnetic flux errors in magnet-moving measurement schemes. The findings demonstrate that selecting an optimal weighing position can reduce the far-end flux effect to the order of $10^{-9}$. In contrast, the near-end flux effect can be quantified by monitoring the magnetic field surrounding the magnet system. In the Tsinghua Kibble balance system, we show that with proper control of external flux sources, {the relative error can be reduced below} \( 1 \times 10^{-8} \) without requiring additional magnetic shielding.
\end{abstract}

\begin{IEEEkeywords}
Kibble balance, magnetic shielding, tabletop instrument, magnetic field measurement, mass metrology.
\end{IEEEkeywords}
\IEEEpeerreviewmaketitle

\section{Introduction}
\label{sec01}
\IEEEPARstart{T}{he} Kibble balance has become one of the most precise instruments for mass realization following the new International System of Units (SI)~\cite{cgpm2018,Kibble1976,Stephan16}. Its measurement process involves two distinct phases: the weighing phase and the velocity phase\cite{li2022irony}. In the weighing phase, a dc-current-carrying coil within a magnetic field generates an ampere force that counterbalances the weight of a test mass, written as
\begin{equation}
    mg=(Bl)_\mathrm{w}I,
    \label{eq:weighing}
\end{equation}
where \( m \) is the mass of the test object, \( g \) is the acceleration due to gravity, \( B \) is the magnetic flux density in the air gap, \( I \) is the current through the coil, and \( l \) is the total length of the coil wire. In the velocity phase, the coil is moved through the same magnetic field, inducing a voltage \( U \) at the coil terminals, which is measured as
\begin{equation}
    U=(Bl)_\mathrm{v}v,
    \label{eq:velo}
\end{equation}
where \( v \) is the velocity of the coil relative to the magnetic field. Ideally, the magnetic-geometrical factor \( Bl \) remains consistent across both phases, i.e. $(Bl)_\mathrm{w}=(Bl)_\mathrm{v}$. By combining (\ref{eq:weighing}) and (\ref{eq:velo}), the $Bl$ factor can be eliminated, yielding the mass of the test object as
\begin{equation}
    m=\frac{UI}{gv}.
\end{equation}

Currently, over a dozen metrology institutes or laboratories, including \cite{NIST,NRC,METAS,NPL3,BIPM,LNE,NIM,KRISS,MSL,UME,PTB,li2023design}, are conducting Kibble balance experiments. A significant trend in this field is the development of tabletop Kibble balance instruments \cite{NPL3,PTB,li2022design,KIBBg1performance,PB-2} to enable broader applications within the mass metrology community. In 2022, Tsinghua University initiated a tabletop Kibble balance project~\cite{li2022design} and has achieved notable progress in recent years~\cite{THUg,THUcurrentsource,THUupdate2024,THUmag,THUupdate2025}. 
The Tsinghua system utilizes a magnet-moving mechanism, and the idea is to address the limited measurement range of the weighing unit and the need to avoid moving the weighing unit during velocity measurements. Similar approaches involving the movement of the magnet system have also been implemented in other experiments, such as the Joule balance at the National Institute of Metrology (NIM, China)~\cite{NIM}, the Kibble balance at the TUBITAK National Metrology Institute (UME, Turkey)~\cite{UME}. However, the magnet-moving scheme introduces a potential measurement error due to the residual field caused by an external magnetic flux, as detailed in Section \ref{sec02}. 
To mitigate this issue, magnetic shielding shells and compensation coils can be employed~\cite{xu2018research,xu2019elimination}. Nevertheless, these solutions often complicate coil operations once the shielding is in place. The UME Kibble balance group proposed to evaluate external field errors through measurements of magnetic field changes with and without the magnetic circuit. Their findings provide valuable information for characterizing and evaluating this effect~\cite{UMEemf}. 

\begin{figure}[tp!]
    \centering
    \includegraphics[width=0.5\textwidth]{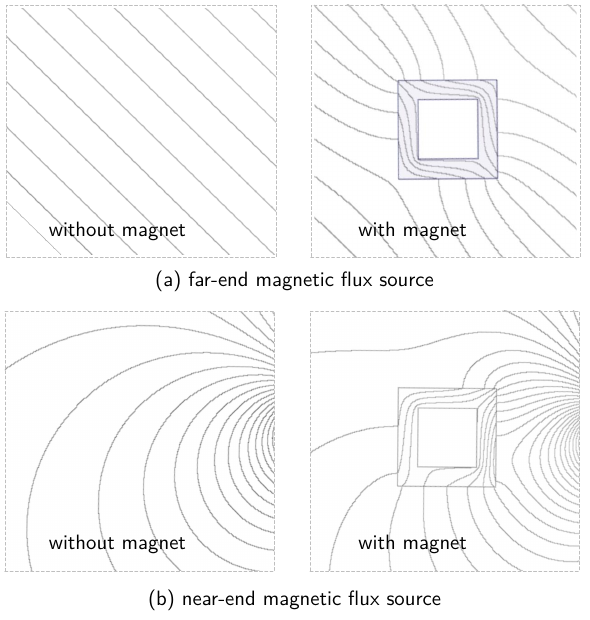}
    \caption{Definition of (a) the far-end magnetic flux and (b) the near-end magnetic flux. The lines in each subplot present the magnetic flux lines, and the region between two rectangles denotes a self-shielded iron (Kibble balance magnet).}
    \label{fig:nearfar}
\end{figure}

The BIPM-type magnetic circuit offers a good self-shielding capability, as the working field is entirely enclosed by iron components\cite{li2022}. It is therefore of interest to quantify the attenuation of external magnetic flux as it passes through the magnet, which can provide insight into the effort required to achieve a systematic effect below \(1 \times 10^{-8}\). As shown in Fig. \ref{fig:nearfar}, external magnetic flux sources can be categorized into two types: (1) far-end flux, which generates a uniform field at the magnet's position, such as the geomagnetic field, and (2) near-end flux, which produces a gradient field at the magnet, for example, the fields generated by nearby electrical equipment. Ref. \cite{BIPMmagShielding} demonstrates that the vertical component of a far-end field can induce a field gradient in the air gap, while its horizontal component generates additional torques in the weighing measurement, which can be mitigated through coil alignment. However, the effect of near-end magnetic fields has not yet been thoroughly investigated.  

In this paper, we examine the impact of external magnetic fields, with a particular focus on near-end fields, and propose a method to evaluate errors in magnet-moving Kibble balances in section \ref{sec02}. Section~\ref{sec03} and~\ref{sec04} analyze the magnetic coupling mechanism of far-end and near-end external fields using finite element analysis (FEA). Section~\ref{sec05} presents experimental measurements conducted with fluxgate magnetic sensors to characterize the near-static magnetic field. In Section~\ref{sec06}, the impact of external magnetic fields is evaluated using the Tsinghua tabletop Kibble balance as a case study. Finally, Section~\ref{sec07} summarizes the findings and conclusions.

\section{External field coupling mechanism and error evaluation method}
\label{sec02}

\subsection{Error related to external magnetic flux}

Without loss of generality, we assume that the external magnetic flux generates a magnetic flux density in the space surrounding the magnet, described in terms of the radial and vertical coordinates \((r, z)\). This external flux passes through the magnetic circuit and induces a radial magnetic field \( \Delta B_r\) at the coil's location. It is important to note that \( \Delta B_r \) is a function of the azimuthal coordinate \( \theta \). 

During the weighing phase of a Kibble balance, the magnetic force acting on the coil is contributed by both the field generated from the magnet \( B \) and the static external flux \( \Delta B_\mathrm{w} \). The corresponding magnetic geometrical factor, represented by the force-to-current ratio, is expressed as
\begin{eqnarray}
(Bl)_\mathrm{w}&=&Bl+\frac{l}{2\pi}\int_0^{2\pi}\Delta B_\mathrm{w}(r_\mathrm{c},\theta,z_\mathrm{w})\mathrm{d}\theta \nonumber\\
&=&Bl+\Delta \bar{B}_\mathrm{w}(z_\mathrm{w})l,
\label{eq:blw}
\end{eqnarray}
where \( r_\mathrm{c} \) is the mean radius of the coil, and \( z_\mathrm{w} \) denotes the vertical position of the weighing point. The term \( \Delta \bar{B}_\mathrm{w}(z) \) represents the average magnetic flux density of \( \Delta B_\mathrm{w} \) along the coil’s circumference at radius \( r_\mathrm{c} \), as a function of the coil’s vertical position \( z \), i.e.
\begin{eqnarray}
    \Delta \bar{B}_\mathrm{w}(z)&=&\frac{1}{2\pi}\int_0^{2\pi}\Delta B_\mathrm{w}(r_\mathrm{c},\theta,z)\mathrm{d}\theta.
    \label{eq:05}
\end{eqnarray}

In the velocity phase, the magnetic geometrical factor \((Bl)_\mathrm{v}\) is determined differently depending on the measurement scheme employed. When a coil-moving mechanism is used, the magnetic factor remains consistent, i.e., \((Bl)_\mathrm{v} = (Bl)_\mathrm{w}\). However, when a magnet-moving mechanism is utilized, the magnetic field generated by the magnet system, represented by \(Bl\), is {observed by the stationary coil}. In this case, a significant portion of the later term in (\ref{eq:blw}) is not captured in the induced voltage measurement, thereby affecting the determination of \((Bl)_\mathrm{v}\). Additional external magnetic flux, such as that generated by the motor driving the magnet, may contribute to the \((Bl)_\mathrm{v}\) measurement in the velocity phase. This additional flux can be expressed similarly to (\ref{eq:05}) as
\begin{eqnarray}
    \Delta \bar{B}_\mathrm{v}(z) &=& \frac{1}{2\pi} \int_0^{2\pi} \Delta B_\mathrm{v}(r_\mathrm{c}, \theta, z) \, \mathrm{d}\theta,
    \label{eq:06}
\end{eqnarray}
where \(\Delta B_\mathrm{v}\) represents the coupled magnetic flux density arising from the newly introduced flux during the velocity measurement, and \(\Delta \bar{B}_\mathrm{v}(z)\) denotes the average magnetic flux density of \(\Delta B_\mathrm{v}\) along the coil's circumference at radius \(r_\mathrm{c}\).

In summary, the external magnetic flux introduces a bias for the mass measurement when the magnet-moving scheme is employed. The error term can be written as
\begin{eqnarray}
\varepsilon &=&\frac{\Delta m}{m}=\frac{B(z_\mathrm{w})+\Delta \bar{B}_\mathrm{w}(z_\mathrm{w})}{B(z_\mathrm{w})+\Delta \bar{B}_\mathrm{v}(z_\mathrm{w})}\nonumber\\
&\approx&\frac{\Delta \bar{B}_\mathrm{w}(z_\mathrm{w})-\Delta \bar{B}_\mathrm{v}(z_\mathrm{w})}{B(z_\mathrm{w})}.
\label{eq:07}
\end{eqnarray}
Note that in (\ref{eq:07}), the term \(\Delta \bar{B}_\mathrm{w}\) arises from the static external magnetic flux, whereas \(\Delta \bar{B}_\mathrm{v}\) originates from newly generated flux sources associated with the velocity measurement. Consequently, in the subsequent discussion, \(\Delta \bar{B}_\mathrm{w}\) and \(\Delta \bar{B}_\mathrm{v}\) are referred to as the static field change component and the dynamic field change component, respectively. 

\subsection{An external flux error evaluation method}

It is important to note that neither \(\Delta \bar{B}_\mathrm{w}\) nor \(\Delta \bar{B}_\mathrm{v}\) in (\ref{eq:07}) is directly measurable. This is because no magnetic sensor currently exists that can simultaneously detect nanotesla (nT)-level signal changes and cover a magnetic field range of sub-Tesla magnitude, e.g., 0.59\,T in the Tsinghua tabletop Kibble balance system~\cite{THUmag}. To overcome this limitation, an indirect approach is proposed to evaluate \(\Delta \bar{B}_\mathrm{w}\) and \(\Delta \bar{B}_\mathrm{v}\). The method consists of two key steps: First, magnetic sensors are placed outside the magnetic circuit to measure the external magnetic field, denoted as \(B_\mathrm{e}\). Second, the field attenuation factor \(\gamma\), which describes the attenuation from \(B_\mathrm{e}\) to the magnetic field in the air gap \(\Delta B_r\), is determined through FEA analysis or experiment. Using these steps, \(\Delta B_r\) can be estimated as
\begin{equation}
    \Delta B_r = B_\mathrm{e} \cdot \gamma.
\end{equation}

The proposal’s principle is straightforward, but its practical implementation involves two key steps that require careful consideration of factors influencing the evaluation.  

In the first step, the external magnetic field \( B_\mathrm{e} \) is weak, and measurements must account for magnetic circuit leakage flux, which can be substantial — often exceeding the external flux being measured. One possible approach is to measure \( B_\mathrm{e} \) in the absence of the magnetic circuit. However, as Fig. \ref{fig:nearfar} illustrates, the yoke alters the external magnetic flux distribution, thereby modifying \( B_\mathrm{e} \). Consequently, the field measured without the magnetic circuit differs from that measured with it.  
A preferable solution is to measure \( B_\mathrm{e} \) with the yoke in place but without the permanent magnets inside the circuit. This approach eliminates magnetic flux leakage from the circuit while preserving the yoke’s reshaping effect on the external magnetic flux distribution.  

In the second step, the evaluation of \(\gamma\) depends on the magnetic flux distribution. For the far-end external flux, \(\gamma\) can be decomposed into its radial (\(r\)) and axial (\(z\)) components. For the near-end external flux, however, spatial field gradients must also be accounted for in the analysis. Finite element analysis (FEA) is generally well-suited for quantifying \(\gamma\). Experimental validation is also feasible, as discussed in Section \ref{eq:05}. A key challenge lies in measuring weak field variations within a strong background field. One solution involves using a magnetic circuit that excludes the permanent magnets, similar as mentioned above. However, in this configuration, \(\gamma\) becomes a function of the yoke’s magnetic permeability (\(\mu\)), necessitating careful consideration of permeability variations.  

In the following sections, we analyze the mass measurement bias in a magnet-moving Kibble balance, using the Tsinghua tabletop system as a case study.

\section{Impact of the far-end external magnetic flux}
\label{sec03}

\begin{figure}
    \centering    
    \includegraphics[width=0.46\textwidth]{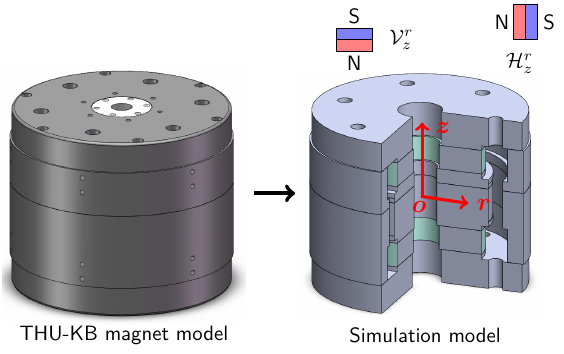}
    \caption{A comparison of the original CAD model and simplified model of the magnet system. The red and blue rectangles represent the north and south poles of a near-end external magnetic flux source used in the near-end field analysis. $\mathcal{V}_z^r$ and $\mathcal{H}_z^r$ denote the vertical and horizontal external magnetic flux sources at the radius $r$ and vertical position $z$.}
    \label{fig:2-1}
\end{figure}

The far-end flux is predominantly geomagnetic in origin, comprising both horizontal and vertical components\cite{Earthmag,Geomag2017}. \cite{BIPMmagShielding} analyzed the impact of the far-end flux in the BIPM-type Kibble balance magnet. This section employs FEA and examines the influence of a far-end external flux for the magnet-moving Kibble balance. The left subplot of Fig. \ref{fig:2-1} presents the 3D model of the Tsinghua magnet system~\cite{THUmag}. To reduce the complexity of the model, components and structures with minimal influence on the magnetic field distribution were excluded from the analysis. The omitted components include screws, aluminum fixing accessories, and similar non-essential parts. Additionally, removed structures primarily consist of screw holes. However, the holes on the cover yoke used for coil connection were retained due to their relatively large diameter of 15\,mm. The resulting simplified model is depicted in the right subplot of Fig. \ref{fig:2-1}. To observe the magnetic field coupled to the air gap, the permanent magnets are replaced by air (permeability $\mu=\mu_0$), and the yoke’s relative magnetic permeability was set to 10 000 (close to the measurement result \cite{THUmag}). The geometrical parameters of the magnet can be found in \cite{THUmag}. The vertical and horizontal components of the far-end magnetic field can be easily determined by measurements. Here we take the geomagnetic field as an example. According to data published by the Institute of Geology and Geophysics, Chinese Academy of Sciences (Beijing, China), the horizontal and vertical components have intensities of approximately 28\,$\upmu$T and 48\,$\upmu$T, respectively~\cite{Geomagnetic}. These signals are used as the external far-end flux source in the calculation.

\begin{figure*}
    \centering
    \includegraphics[width=0.85\textwidth]{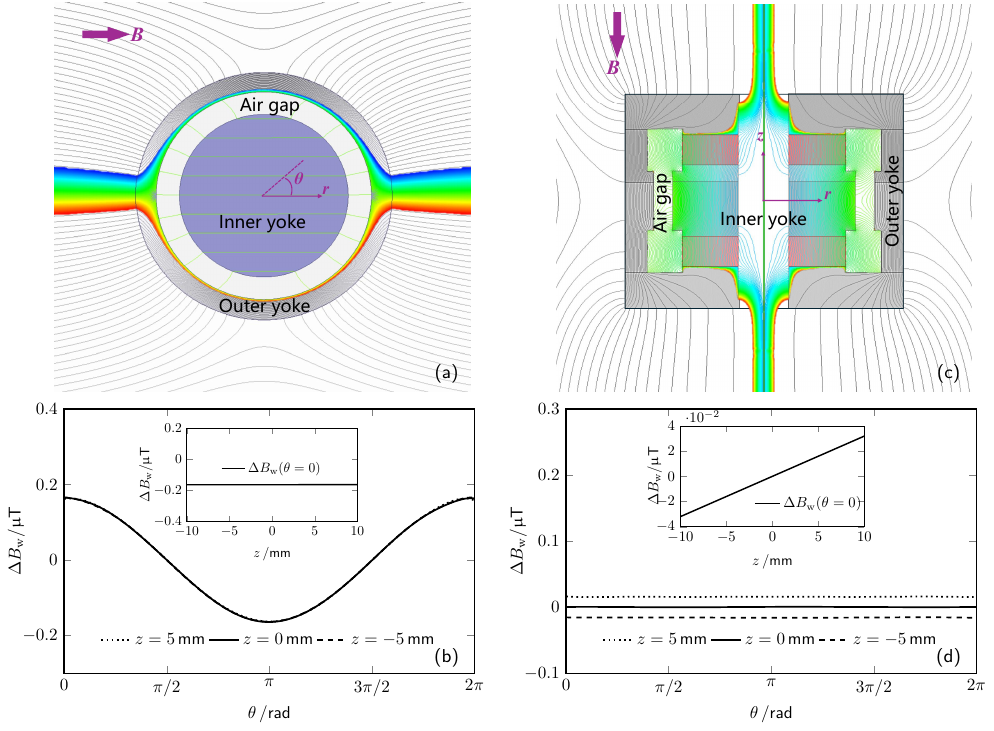}
    \caption{(a) and (c) illustrate the coupling path of far-end magnetic flux in the horizontal and vertical directions, respectively. The flux lines are represented by black curves, while the colored regions highlight detailed flux line distributions through the air gap. (b) displays the radial magnetic field generated within the magnet system by the horizontal far-end flux component. The main panel shows the angular distribution, with the inset presenting the axial ($z$-direction) distribution at $\theta=0$. (d) presents the coupled magnetic field distribution resulting from far-end vertical magnetic excitation. Similar to (b), the main panel depicts angular variation, while the inset shows the axial profile at $\theta=0$.}
    \label{fig:GM}
\end{figure*}

\subsection{Horizontal component}

The path of the horizontal far-end magnetic field entering the magnet system is illustrated in Fig. \ref{fig:GM}(a). The yoke reshapes the magnetic flux distribution, and the radial magnetic field generated by the horizontal component, $\Delta \bar{B}_{\mathrm{w}}$, exhibits a good symmetric angular distribution, as seen in Fig. \ref{fig:GM}(b), where the inward direction (toward the magnet system) is defined as positive and the outward direction as negative. Owing to the yoke’s shielding effect, the horizontal geomagnetic field attenuates at least by a factor of 170 (peak value). Because the horizontal component is uniform along the vertical direction and the yoke thickness around the air gap is constant, the \(z\)-axis distribution remains uniform, as shown in the inset of Fig. \ref{fig:GM}(b).   
Based on the radial magnetic field distribution induced by the horizontal component, the coupled magnetic field during the weighing phase is given by 
\begin{equation}  
    \Delta \bar{B}_{\mathrm{w-FH}} = \int_{0}^{2\pi} \Delta \bar{B}_{\mathrm{w}}(r_\mathrm{c},\theta,z_\mathrm{w}) \, \mathrm{d}\theta = 0.  
    \label{eq:FH}  
\end{equation}  
As (\ref{eq:FH}) indicates, the net additional force is zero, but a flipping torque is generated when a current passes through the coil.  

\subsection{Vertical component}

The coupling path of the vertical component is illustrated in Fig. \ref{fig:GM}(c). A portion of the magnetic flux entering the yoke generates a radial magnetic field as it traverses the air gap in the magnet system. Furthermore, analysis of the coupling path indicates that the radial magnetic fields on either side of the horizontal symmetry plane are oppositely directed. The corresponding simulation results are presented in Fig. \ref{fig:GM}(d).  
The yoke exhibits an outstanding shielding effect on the vertical component, reducing the internal magnetic field to $\pm 16$\,nT at $z=\pm 5$mm -- a suppression factor of approximately 300 relative to the external geomagnetic field. While the angular distribution remains uniform, $\Delta B_r$ shows a linear dependence on $z$, as demonstrated in the inset of Fig. \ref{fig:GM}(d). In this configuration, the average coupled magnetic field in the air gap, denoted as $\Delta \bar{B}_{\mathrm{w-FV}}$, exhibits a linear dependence on the vertical position $z_\mathrm{w}$. This relationship can be expressed as:  
\begin{equation}  
    \Delta \bar{B}_{\mathrm{w-FV}} = \int_{0}^{2\pi} \Delta \bar{B}_{\mathrm{w}}(r_\mathrm{c},\theta,z_\mathrm{w}) \, \mathrm{d}\theta = \alpha z_{\mathrm{w}},  
    \label{eq:FV}  
\end{equation}  
where $\alpha = 3.2$\,nT/mm represents the magnetic coupling coefficient.  
Eq. (\ref{eq:FV}) reveals that the resulting measurement bias is directly proportional to the weighing position $z_\mathrm{w}$. Optimal suppression of this effect is achieved when weighing at the symmetry plane ($z_\mathrm{w} = 0$). For the Tsinghua Kibble balance system, every vertical displacement of 1\,mm from the symmetry plane induces a relative bias of 3.2\,nT/0.59\,T$\approx 5.4 \times 10^{-9}$.

\section{FEA analysis of the near-end magnetic flux}
\label{sec04}

\subsection{Simulation setup}

As discussed in Section~\ref{sec02}, evaluating the measurement error caused by near-end magnetic flux requires an investigation of the attenuation factor, $\gamma$, which quantifies how much the external flux decays when passing through the yoke. Here, we perform 3D FEA simulations to explore how different external flux sources interact with the magnet system. Permanent magnets are selected as the source of near-static magnetic interference for both simulation and experimental studies. This choice is motivated by their ability to generate stable interfering magnetic fields, whose characteristics can be easily adjusted by modifying the position and magnetization direction of the magnets.  

Prior to examining the simulation details, we establish the spatial coordinate system and classify external magnetic sources according to the magnetization direction of interfering permanent magnets, as illustrated in Fig. \ref{fig:2-1}. The coordinate system is defined with the following conventions: The vertical symmetry axis of the magnet system serves as the $z$-axis. The origin is located at the intersection of the horizontal symmetry plane and the $z$-axis. We categorize magnetic interference sources as: (1) horizontal near-end external flux ($\mathcal{H}_z^r$) where permanent magnet is magnetized in the horizontal direction and (2) vertical near-end external flux ($\mathcal{V}_z^r$), where the permanent is magnetized in the vertical direction. Note that the superscript $r$ and the subscript $z$ denote respectively the radial and vertical position of the near-end magnetic flux. 

In the simulation, a cylinder permanent magnet (diameter $\phi=30$\,mm and height $h=10$\,mm) is employed. The residual coercivity of the interfering magnet is set to $-820\,$kA/m, and in
this case the surface magnetic field strength $B_{\rm s}$ of the interfering magnet is 0.336\,T.

\subsection{FEA results}
\label{sec2-2}

The outer radius of the Tsinghua Kibble balance magnet is 110\,mm, and the height is 180\,mm ($z$ ranges from -90\,mm to 90\,mm). In the following, we show the FEA results when the interference permanent magnet is placed at several typical positions around the magnet circuit.

\begin{figure*}[tp!]
    \centering
    \includegraphics[width=0.9\textwidth]{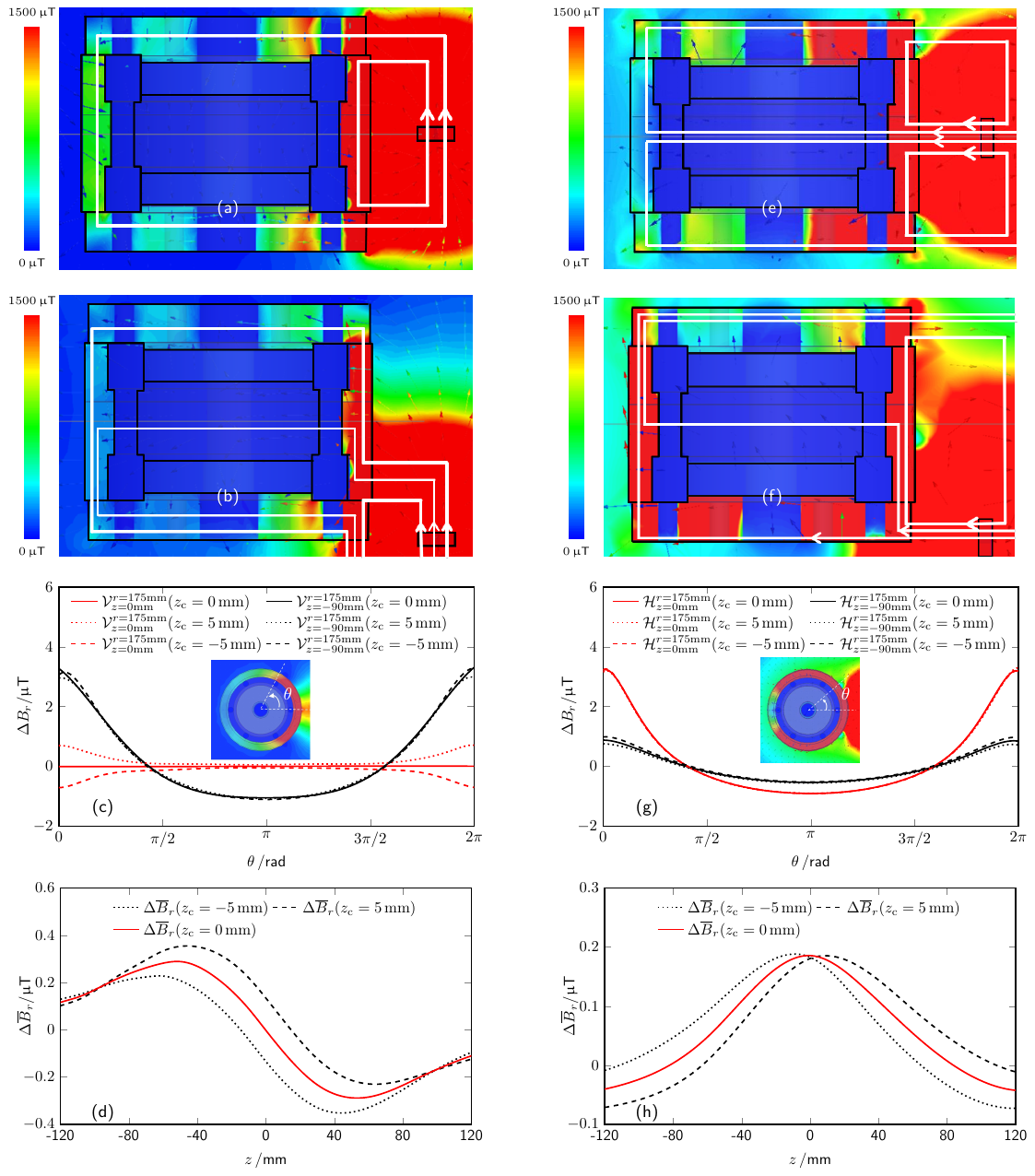}
    
    \caption{Calculation results of the $\mathcal{V}^{r=175\,\mathrm{mm}}_{z}$ and $\mathcal{H}^{r=175\,\mathrm{mm}}_{z}$. (a) and (b) are the magnetic flux density cloud maps of $\mathcal{V}^{r=175\,\mathrm{mm}}_{z}$ on the $zy$ plane when $z=0\,$mm and $z=-90\,$mm, respectively. The white curves represent the main coupling path of the external magnetic flux. (c) is the angular distribution of the air-gap magnetic field generated by $\mathcal{V}^{r=175\,\mathrm{mm}}_{z}$, and the subplot is the cloud map of the magnetic flux density on the plane of $z=0$\,mm. (d) is the average filed of $\mathcal{V}^{r=175\,\mathrm{mm}}_{z}$ at various $z$-position.
     (e) and (f) are the magnetic flux density cloud maps of the $\mathcal{H}^{r=175\,\mathrm{mm}}_{z}$ on the $zy$ plane when $z=0\,$mm and $z=-90\,$mm, respectively. (g) is the angle-distribution of the air-gap magnetic field generated by $\mathcal{H}^{r=175\,\mathrm{mm}}_{z}$, and the subplot is the cloud map of the magnetic flux density on the plane of $z=0$\,mm. (h) is the average filed of $\mathcal{H}^{r=175\,\mathrm{mm}}_{z}$ at various $z$ positions. }
    \label{fig:4}
\end{figure*}

\begin{enumerate}
    \item $\mathcal{V}^{r>110\mathrm{mm}}_{z}$: 
    In this configuration, the external-source magnet is positioned adjacent to the Kibble balance magnet, with its magnetic poles aligned along the $z$-axis. Fig. \ref{fig:4}(a) and (b) present the magnetic field distributions for two cases: $\mathcal{V}^{r=175\,\mathrm{mm}}_{z=0\,\mathrm{mm}}$ and $\mathcal{V}^{r=175\,\mathrm{mm}}_{z=-90\,\mathrm{mm}}$. The yoke enclosure effectively guides most of the external magnetic flux back to the external permanent magnet, substantially reducing the coupled magnetic field in the air gap. When the external magnet is aligned with the symmetry plane ($z=0\,\mathrm{mm}$), the theoretical magnetic field at the air gap center is zero. The corresponding radial magnetic field distribution along the coil's circular path, $\theta \in (0, 2\pi)$, is depicted by the solid red curve in Fig. \ref{fig:4}(c). However, when the external magnet is displaced above or below $z=0\,\mathrm{mm}$, flux penetrates the air gap, resulting in a non-uniform $\Delta B_r(\theta)$ distribution. Peak values appears at $\theta = 0$ (3.3\,$\upmu$T), and $\theta = \pi$ (-1.1\,$\upmu$T). The maximum value of the attenuate factor (${\gamma}_{\mathrm{max}}$) is $9.8 \times 10^{-6}$.
    Changing the coil position $z_\mathrm{c}$ can slightly shift the $\Delta B_r(\theta)$ distribution, as can be seen two additional cases, $z_\mathrm{c}=\pm5$\,mm, presented in \ref{fig:4}(c). Fig. \ref{fig:4} (d) shows the change of the average of $\Delta B_r$ as a function of the vertical position $z$ of the external permanent magnet. With $z_\mathrm{c}=0$\,mm, the maximum value of the averaged magnetic field is 0.3\,$\upmu$T at $z=\pm 51$\,mm. Compared to the external magnetic field at the permanent magnet (336\,mT), the average attenuation factor ${\gamma}_{\mathrm{ave}}\approx8.6\times 10^{-7}$. 

    \item $\mathcal{H}^{r>110\mathrm{mm}}_{z}$: 
    In this case, the magnetic poles of the external magnet aligned along the $r$-axis. {Fig. \ref{fig:4}(e)} and (f) present the magnetic field distributions for two cases: $\mathcal{H}^{r=175\,\mathrm{mm}}_{z=0\,\mathrm{mm}}$ and $\mathcal{H}^{r=175\,\mathrm{mm}}_{z=-90\,\mathrm{mm}}$. Compared to $\mathcal{H}^{r=175\,\mathrm{mm}}_{z=0\,\mathrm{mm}}$, the average magnetic flux density within the outer yoke with $\mathcal{H}^{r=175\,\mathrm{mm}}_{z=-90\,\mathrm{mm}}$ is higher, which means less external magnetic flux of $\mathcal{H}^{r=175\,\mathrm{mm}}_{z=-90\,\mathrm{mm}}$ can pass through the air gap and return to the external magnet. 
    Hence, in theory, the $\Delta B_r$ of $\mathcal{H}^{r=175\,\mathrm{mm}}_{z=0\,\mathrm{mm}}$ is larger than that of $\mathcal{H}^{r=175\,\mathrm{mm}}_{z=-90\,\mathrm{mm}}$.
    Fig. \ref{fig:4}(g) displays the field distributions of $\mathcal{H}^{r=175\,\mathrm{mm}}_{z=0\,\mathrm{mm}}$ and $\mathcal{H}^{r=175\,\mathrm{mm}}_{z=-90\,\mathrm{mm}}$. When the external magnet is located at the central symmetrical plane, $z=0$\,mm, its magnetic flux has to pass through the air gap, creating a peak value at $\theta=0$ and a valley at $\theta=\pi$. Moving the external magnet either up or down, less external flux passes through the central plane $z=0$\,mm and hence produces a lower coupled field at the coil position. 
    As predicted, the peak of the red curves ($\mathcal{H}^{r=175\,\mathrm{mm}}_{z=0\,\mathrm{mm}}$), $\approx3.2\,\upmu$T, is 2.5 times higher than the black curves ($\mathcal{H}^{r=175\,\mathrm{mm}}_{z=-90\,\mathrm{mm}}$), 0.9\,$\upmu$T.
    The maximum value of $\gamma$ is $9.6 \times 10^{-6}$. Fig. \ref{fig:4} (h) presents the change of the average of $\Delta B_r$ with $\mathcal{H}^{r=175\,\mathrm{mm}}_{z}$ at various $z$-coordinates. The peak of $\Delta \overline{B}_{r}$ is 0.185\,$\upmu$T, with the external magnet at $z=0$\,mm, and the average attenuation factor is $5.5 \times 10^{-7}$.

\begin{figure*}[t]
    \centering
    \includegraphics[width=0.9\textwidth]{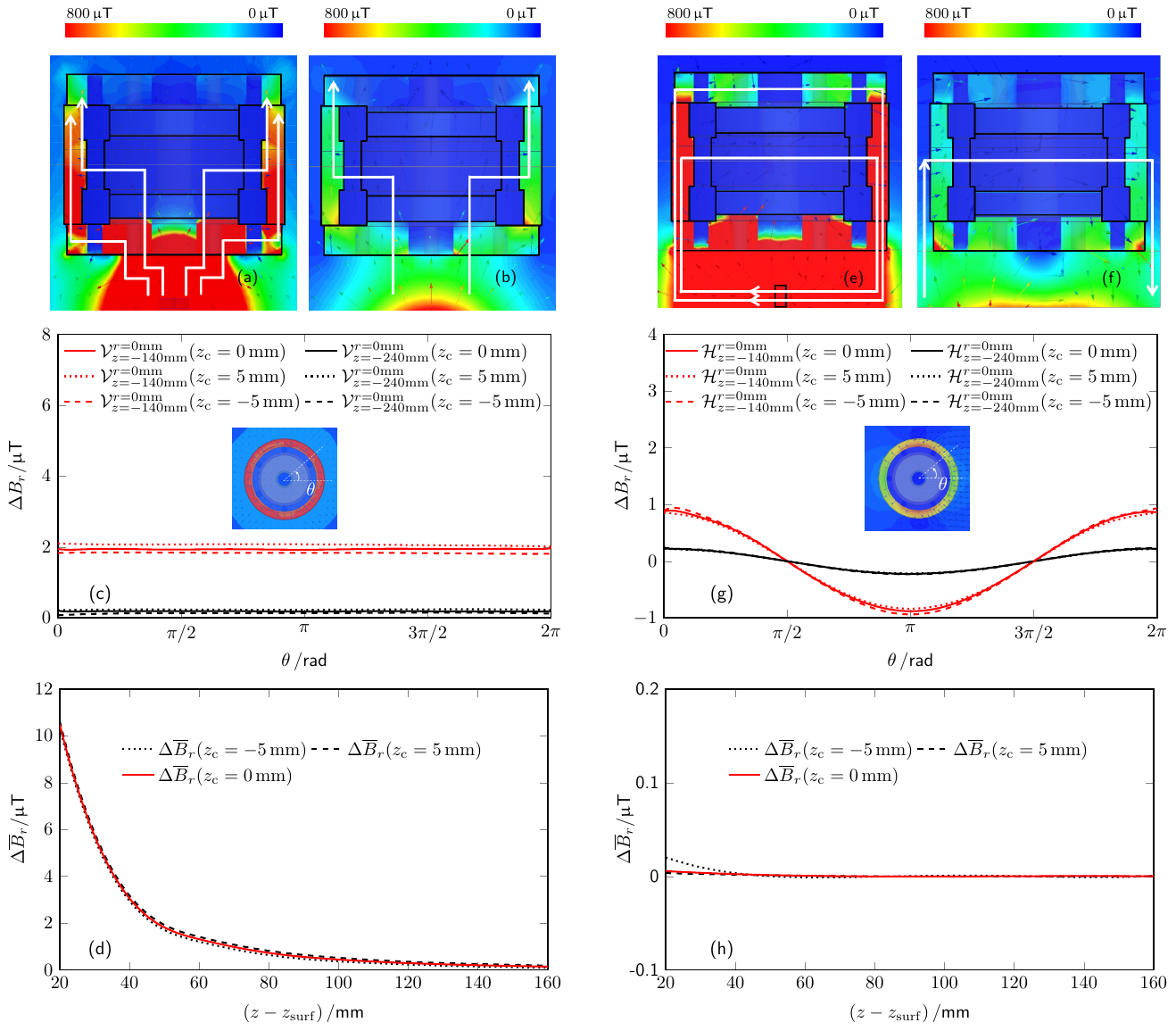}
    \caption{Calculation results of the $\mathcal{H}^{r=0\,\mathrm{mm}}_{z<-90\,\mathrm{mm}}$ and $\mathcal{V}^{r=0\,\mathrm{mm}}_{z<-90\,\mathrm{mm}}$. 
    (a) and (b) are the magnetic flux density cloud maps of the $\mathcal{V}^{r=0\,\mathrm{mm}}_{z}$ on the $zy$ plane when $z=-140\,$mm and $z=-190\,$mm, respectively. (c) is the angular distribution of the air-gap magnetic field generated by $\mathcal{V}^{r=0\,\mathrm{mm}}_{z<-90\,\mathrm{mm}}$, and the subplot is the cloud map of the magnetic flux density on the plane of $z=0$\,mm. (d) is the average field of $\mathcal{H}^{r=0\,\mathrm{mm}}_{z}$.
    (e) and (f) are the magnetic flux density cloud maps of $\mathcal{H}^{r=0\,\mathrm{mm}}_{z}$ on the $zy$ plane when $z=-140\,$mm and $z=-190\,$mm, respectively. The white curves represent the main coupling path of the external magnetic flux. (g) is the angle-distribution of the air-gap magnetic field generated by $\mathcal{H}^{r=0\,\mathrm{mm}}_{z<-90\,\mathrm{mm}}$, and the subplot is the cloud map of the magnetic flux density on the plane of $z=0$\,mm. (h) is the average field of $\mathcal{H}^{r=0\,\mathrm{mm}}_{z}$. $z_\mathrm{surf}$ is the top surface of the magnet circuit (top yoke cover).}
    \label{fig:5}
\end{figure*}

     \item $\mathcal{V}^{r=0\mathrm{mm}}_{z<-90\mathrm{mm}}$: 
    In this case, the magnetic poles of the external magnet aligned along the $z$-axis. Fig. \ref{fig:5}(a) and (b) show the magnetic field distributions of $\mathcal{V}^{r=0\,\mathrm{mm}}_{z=-140\,\mathrm{mm}}$ and $\mathcal{V}^{r=0\,\mathrm{mm}}_{z=-240\,\mathrm{mm}}$.
    Due to the symmetry of the system, the magnetic flux lines uniformly diverge towards the outer yoke from the inner yoke, yielding a uniform distribution of $\Delta B_r(\theta)$. Fig. \ref{fig:5}(c) is the angle distribution of the coupled magnetic field in the air gap, and as analyzed, $\Delta B_r$ is uniform along $\theta$. The value is 1.9\,$\upmu$T for $\mathcal{V}^{r=0\,\mathrm{mm}}_{z=-140\,\mathrm{mm}}$ and 0.2\,$\upmu$T for $\mathcal{V}^{r=0\,\mathrm{mm}}_{z=-240\,\mathrm{mm}}$. 
    Fig. \ref{fig:5}(d) displays the change of $\Delta \overline{B}_{r}$ with the change of the external magnet's $z$-coordinate. The coupled magnetic field has a rapid increase when the external magnet gets closer to the magnetic circuit. As the distance increases, the air-gap magnetic field is gradually approaching zero.
 
    \item $\mathcal{H}^{r=0\mathrm{mm}}_{z<-90\mathrm{mm}}$:
    In this case, the magnetic poles of the external magnet are aligned along the $r$-axis, and its geometric center is on the symmetry axis $r=0$\,mm. Fig. \ref{fig:5}(e) and (f) present the magnetic field distributions of two vertical positions, i.e. $\mathcal{H}^{r=0\,\mathrm{mm}}_{z=-140\,\mathrm{mm}}$ and $\mathcal{H}^{r=0\,\mathrm{mm}}_{z=-240\,\mathrm{mm}}$, respectively.
    The flux path contains a symmetry in this case, i.e. the external magnetic flux enters from one side of the magnet circuit and exits from the opposite side. Although a larger $\Delta B_r(\theta)$ variation of the coupled magnetic field in the air gap will be obtained when the external magnet stays closer to the magnetic circuit, the average field $\Delta \bar{B}$ should remain zero. As shown in Fig. \ref{fig:5}(g), the air-gap magnetic field's peak value, $\pm 0.9\,\upmu$T for $\mathcal{H}^{r=0\,\mathrm{mm}}_{z=-140\,\mathrm{mm}}$ and $\pm 0.2\,\upmu$T for $\mathcal{H}^{r=0\,\mathrm{mm}}_{z=-240\,\mathrm{mm}}$, appears on $\theta = 0$ and $\pi$. Here, the maximum value of the attenuation factor is $2.5 \times 10^{-6}$. Fig. \ref{fig:5} (h) presents the change of $\Delta \overline{B}_{r}$ with the change of the external magnet's position. Due to the symmetry, $\Delta \overline{B}_{r}$ equals zero.
\end{enumerate}

\subsection{Near-end external field limitation}
\label{Sec:2-3}

As discussed in Section~\ref{sec2-2}, the inhomogeneous magnetic field induced by the external source in the air gap introduces a discrepancy between $(Bl)_{\mathrm{w}}$ and $(Bl)_{\mathrm{v}}$, resulting in a systematic bias in the mass measurement. To minimize this effect, the near-end magnetic flux must be carefully controlled. Using the FEA results, the influence of the external magnetic field can be quantitatively constrained.
To ensure a measurement error below $1 \times 10^{-8}$, the external magnetic flux density $B_{\mathrm{e}}$ must satisfy
\begin{equation}
    B_{\mathrm{e}} \leq \frac{B}{\gamma} \times 10^{-8},
\end{equation}
where $B\approx0.59\,\mathrm{T}$ is the main magnetic field generated by the internal permanent magnets~\cite{THUmag}. To be safe, the maximum attenuation factor $\ \gamma_{\mathrm{max}}$ is chosen to calculate the limitation of external fluxes.

We analyze the four cases introduced in Section~\ref{sec2-2}, i.e. $\mathcal{V}^{r>110\,\mathrm{mm}}_{z=-90\,\mathrm{mm}}$, $\mathcal{H}^{r>110\,\mathrm{mm}}_{z=0\,\mathrm{mm}}$, $\mathcal{V}^{r=0\,\mathrm{mm}}_{z<-90\,\mathrm{mm}}$, and $\mathcal{H}^{r=0\,\mathrm{mm}}_{z<-90\,\mathrm{mm}}$. For each configuration, we consider three representative locations where the external permanent magnet is placed at distances ($D$) of 50\,mm, 150\,mm, and 250\,mm from the outer surface of the magnet system. 
The corresponding limiting external magnetic fields ${(B_{\rm e})}_{\rm max}$ are summarized in Table~\ref{tab:1}. The results reveal that the attenuation factors at 150\,mm and 250\,mm are nearly identical, suggesting that the attenuation factor becomes effectively distance-independent when the interference source is sufficiently far from the magnetic circuit.
 
\begin{table}
     \centering
     \caption{The minimum attenuation rate and maximum amplitude of the interfering magnetic field.}
     \renewcommand{\arraystretch}{1.8}
     \begin{tabular}{>{\centering\arraybackslash}p{0.1\textwidth} >{\centering\arraybackslash}p{0.08\textwidth} >{\centering\arraybackslash}p{0.1\textwidth} >{\centering\arraybackslash}p{0.1\textwidth}}
     \hline
     \hline
     Type & $D$\,/mm & ${\gamma}_{\rm max}$ & ${(B_{\rm e})}_{\rm max}$\,/mT \\
     \hline

     $\mathcal{V}^{r=160\,\mathrm{mm}}_{z=-90\,\mathrm{mm}}$ & 50 & $9.22\times 10^{-6}$ &0.64 \\
     $\mathcal{V}^{r=260\,\mathrm{mm}}_{z=-90\,\mathrm{mm}}$ & 150 &$1.77\times 10^{-6}$ &3.33 \\
     $\mathcal{V}^{r=360\,\mathrm{mm}}_{z=-90\,\mathrm{mm}}$ & 250 &$4.89\times 10^{-7}$ &12.07 \\
      \hline
     $\mathcal{H}^{r=160\,\mathrm{mm}}_{z=0\,\mathrm{mm}}$ & 50 & $9.51\times 10^{-6}$ & 0.62 \\
     $\mathcal{H}^{r=260\,\mathrm{mm}}_{z=0\,\mathrm{mm}}$ & 150 &$1.61\times 10^{-6}$ &3.66 \\
     $\mathcal{H}^{r=360\,\mathrm{mm}}_{z=0\,\mathrm{mm}}$ & 250 &$5.56\times 10^{-7}$ &10.61 \\
      \hline

     $\mathcal{V}^{r=0\,\mathrm{mm}}_{z=-140\,\mathrm{mm}}$ & 50 & $5.83\times 10^{-6}$ &1.01 \\
     $\mathcal{V}^{r=0\,\mathrm{mm}}_{z=-240\,\mathrm{mm}}$ & 150 &$5.74\times 10^{-7}$ &10.28 \\
     $\mathcal{V}^{r=0\,\mathrm{mm}}_{z=-340\,\mathrm{mm}}$ & 250 &$1.29\times 10^{-7}$ &45.74 \\
      \hline  
     $\mathcal{H}^{r=0\,\mathrm{mm}}_{z=-140\,\mathrm{mm}}$ & 50 & $2.65\times 10^{-6}$ &2.23 \\
     $\mathcal{H}^{r=0\,\mathrm{mm}}_{z=-240\,\mathrm{mm}}$ & 150 &$6.62\times 10^{-7}$ &8.91 \\
     $\mathcal{H}^{r=0\,\mathrm{mm}}_{z=-340\,\mathrm{mm}}$ & 250 &$2.44\times 10^{-7}$ &24.18 \\
      \hline
         
    \hline
     \end{tabular}
     \label{tab:1}
 \end{table}

\section{An Experimental Check on FEA Results}
\label{sec05}

\subsection{Experimental setup}

\begin{figure}[t]
    \centering
    \includegraphics[width=0.45\textwidth]{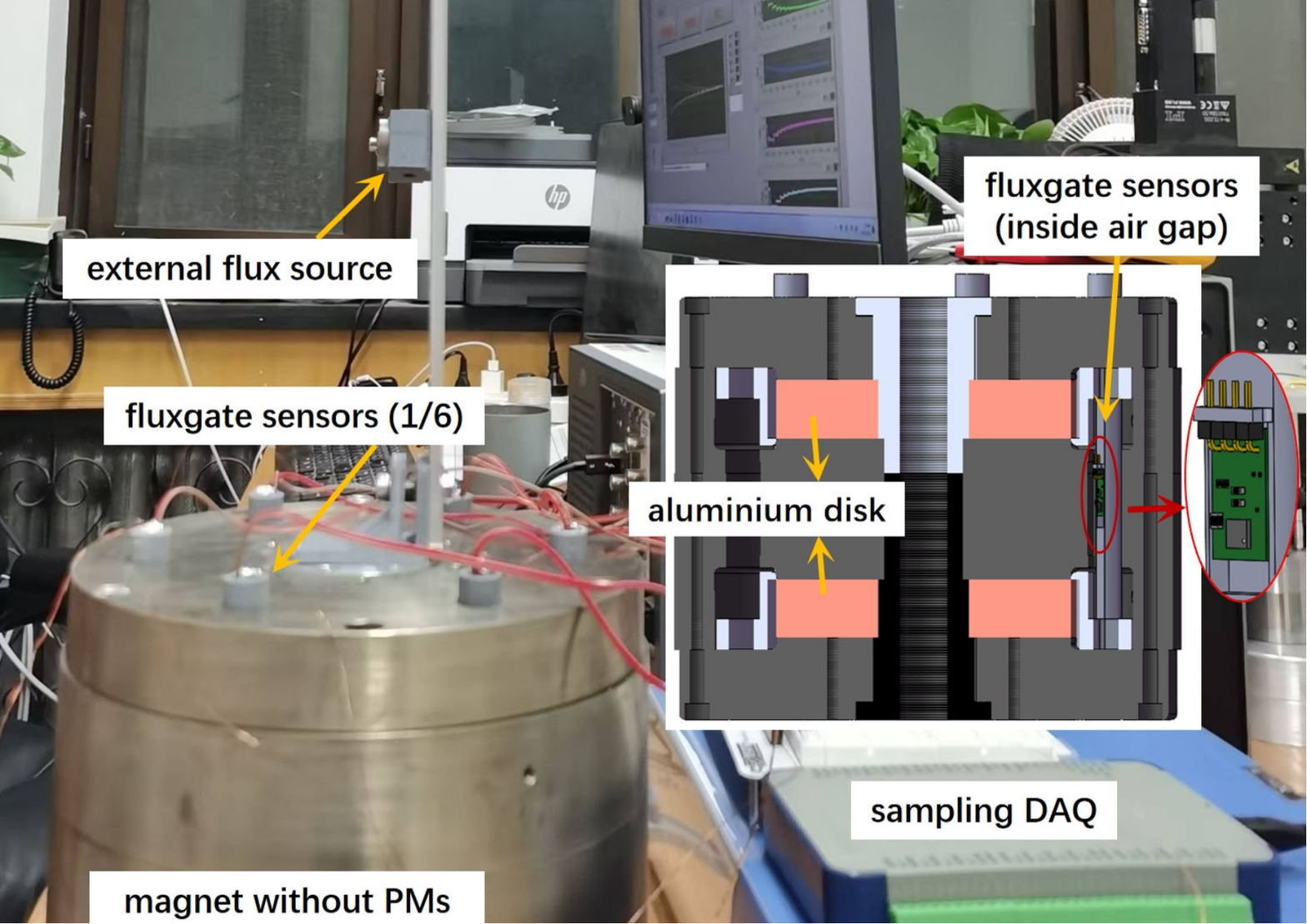}
    \caption{Experimental setup of the measurement. The subplot presents a sectional view of the magnet. In the experiment, the permanent magnets (PMs) were replaced by two aluminum blocks of equal size. Six fluxgate sensors were placed on the horizontal plane of symmetry to measure $\Delta B_r$.}
    \label{fig:experimental_setup}
\end{figure}

 This section presents an experimental check on the FEA calculation in section \ref{sec04}. The experimental measurement system is illustrated in Fig.~\ref{fig:experimental_setup}. As the permanent magnets in the magnetic circuit typically generate a sub-tesla magnetic field in the air gap, while external interfering fluxes produce much weaker fields (e.g., on the order of $\upmu$T), no commercial magnetic sensors exist to achieve sufficient sensitivity across such a wide dynamic range. To mitigate this challenge, the experiment replaced the permanent magnets with two aluminum blocks. In such a way, the field change caused by the external magnetic flux is measurable using weak field sensors.

 A samarium-cobalt magnet was positioned at various locations to generate different external near-field magnetic flux. The magnet has an inner diameter of 4\,mm, an outer diameter of 25\,mm, and a height of 7.5\,mm. The surface magnetic flux density at the center of the interfering magnet, measured using a Gauss meter (Lakeshore Model\,425), is 260\,mT. The coupled magnetic field within the magnet system is on the order of tens of microteslas, necessitating the use of weak magnetic sensors with sufficient sensitivity.  
 As shown in Fig.~\ref{fig:experimental_setup}, six fluxgate sensors (DRV425EVM) are mounted on 3D-printed plastic brackets and inserted into the center of the air gap through six symmetrical holes. The sensors are placed on the same horizontal plane and oriented to measure the radial component of the magnetic field. The output voltage \( U_{\rm out} \) is linearly proportional to the magnetic flux density \( B \), with a conversion factor of 205\,$\upmu$T/V.
 A data acquisition card (DAQ) is employed to collect the output signals from the six fluxgate sensors. The sensor outputs include inherent offsets caused by ambient fields such as the geomagnetic field. To isolate the contribution from background magnetic flux, the offset of each channel is first measured under zero-field conditions (no active external magnetic excitation). By subtracting these baseline offsets, the processed sensor readouts can represent only the coupled field induced by the external magnetic flux.

\subsection{Experimental results}

\begin{figure*}[t]
    \centering
    \includegraphics[width=0.41\textwidth]{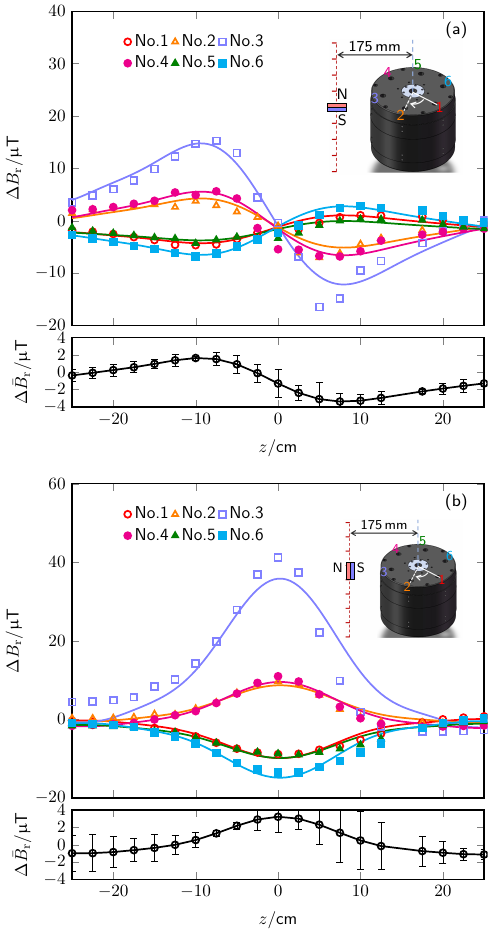}
    \includegraphics[width=0.41\textwidth]{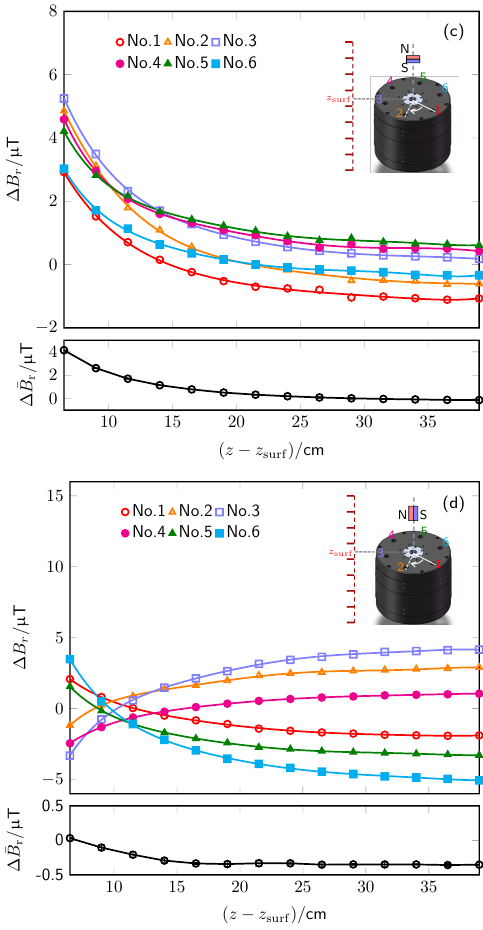}
    \caption{Measurement results of the coupled magnetic field in the air gap excited by the external-source magnet. 
    (a) $\mathcal{V}^{r=175\,\mathrm{mm}}_{z}$, (b) $\mathcal{H}^{r=175\,\mathrm{mm}}_{z}$, (c) $\mathcal{V}^{r=0\,\mathrm{mm}}_{z}$ and (d) $\mathcal{H}^{r=0\,\mathrm{mm}}_{z}$. In each figure, the upper subplot gives the angular distribution and the variation of $\Delta B_r$ as the external magnet's position changes, and the lower subplot presents the $\Delta \overline{B}_r(z)$ function curve. $z_\mathrm{surf}$ is the top surface of the magnet circuit (top yoke cover).
    }
    \label{fig:7}
\end{figure*}

Several measurements similar to examples in {Section \ref{sec04}} were carried out in the experiment: 
(a) $\mathcal{V}^{r=175\,\mathrm{mm}}_{z}$, 
(b) $\mathcal{H}^{r=175\,\mathrm{mm}}_{z}$, 
(c) $\mathcal{V}^{r=0\,\mathrm{mm}}_{z}$ and
(d) $\mathcal{H}^{r=0\,\mathrm{mm}}_{z}$. 
The coupled radial magnetic field of six sensors of $\mathcal{V}^{r=175\,\mathrm{mm}}_{z}$ as a function of the vertical position of the external permanent magnet is presented in the top subplot of Fig. \ref{fig:7} (a). The measurement results are consistent with the FEA results: The peak values occur at $\theta =0$ (No. 3 sensor) and $\theta =\uppi$ (No.6 sensor). At $z_\mathrm{c}=0\,$mm, the coupled magnetic field is zero due to the symmetry. With $z_\mathrm{c}$ getting far away from the symmetrical plane, the coupled field first reaches the peak value and then decays to zero. The peak value is about 15\,$\rm{\upmu T}$ at $z_\mathrm{c}=\pm 75$\,mm. 

 It is observed that $\Delta B_r(z)$ of different sensors has a similar shape of distribution. Here we extracted a shape function by normalizing the amplitude of each sensor’s readout and the shape function is then employed to fit the measurement results, as presented by the solid curves in Fig. \ref{fig:7} (a). The average of the fit gives the change on the $\Delta \overline{B}_r$  as shown in the lower subplot of Fig. \ref{fig:7} (a). 
 The averaged field distribution has a peak-peak value of 2.5\,$\upmu$T. An offset of 0.85\,$\upmu$T is observed, which is likely a result of the imperfect average of six sensors. 
 The $\Delta B_{r}$ curve of the $\mathcal{V}^{r=175\,\mathrm{mm}}_{z}$ is consistent with the simulation result in Fig. \ref{fig:4} (a).
 The measurement yields ${\gamma}_{\rm max}\approx 5.9 \times 10^{-5}$ and ${\gamma}_{\mathrm{ave}}\approx 9.6\times 10^{-6}$.

 The measurement result of \(\mathcal{H}^{r=175\,\mathrm{mm}}_{z}\) is presented in Fig.~\ref{fig:7}(b). Similar to \(\mathcal{V}^{r=175\,\mathrm{mm}}_{z}\), Sensor~No.~3 (\(\theta = 0\)) measures the maximum \(\Delta B_{r}\), while Sensor~No.~6 (\(\theta = \uppi\)) yields the minimum peak value. Using a normalized shape function, the peak values for Sensors~No.~3 and No.~6 are \(35.8\,\upmu\mathrm{T}\) and \(-14.9\,\upmu\mathrm{T}\), respectively, at \(z = 0\,\mathrm{mm}\). The maximum normalized deviation, \(\gamma_{\mathrm{max}}\), is approximately \(1.4 \times 10^{-4}\).  
 The upper subplot shows that the coupled air-gap magnetic field is strongest at the horizontal symmetry plane (\(z = 0\,\mathrm{mm}\)). As the external magnet moves away from this plane, the additional field weakens.  
 The lower subplot displays the six-sensor average, \(\Delta \overline{B}_{r}\). Its distribution matches the FEA result in Fig.~\ref{fig:4}(h), with a peak value of \(3.2\,\upmu\mathrm{T}\) at \(z = 0\,\mathrm{mm}\), giving \(\gamma_{\mathrm{ave}} \approx 1.2 \times 10^{-5}\).  

 Fig.~\ref{fig:7}(c) presents the experimental results for \(\mathcal{V}^{r=0\,\mathrm{mm}}_{z}\). The upper subplot displays \(\Delta B_r\) measurements from the six sensors, which show close agreement and are consistent with theoretical predictions. The lower subplot shows the averaged field \(\Delta \overline{B}_r\). Both \(\Delta B_r\) and \(\Delta \overline{B}_r\) exhibit a gradual decrease as the external magnet moves away from the reference position, matching the trend observed in simulations. The maximum values occur at \(z = 155\,\mathrm{mm}\), with \(\Delta B_r\) reaching \(5.2\,\upmu\mathrm{T}\) and \(\Delta \overline{B}_r\) peaking at \(4.1\,\upmu\mathrm{T}\). Consequently, the attenuation factors are \(\gamma_{\mathrm{max}} = 2.0 \times 10^{-5}\) and \(\gamma_{\mathrm{ave}} = 1.6 \times 10^{-5}\).
 However, it is evident that for each sensor \(\Delta B_r\) does not converge to zero but instead stabilizes at a residual value. This phenomenon can likely be attributed to residual magnetization in the yoke, i.e., residual magnetism persists in the yoke even after the external magnetic source is removed. Notably, this drift effect was not observed in the \(\mathcal{V}^{r=175\,\mathrm{mm}}_{z}\) and \(\mathcal{H}^{r=175\,\mathrm{mm}}_{z}\) experiments. This absence can be explained by the varying magnetization orientation of the yoke at different external magnet positions, which prevented the accumulation of a considerable magnetization.  

 The \(\mathcal{H}^{r=0\,\mathrm{mm}}_{z}\) configuration is positioned above the magnet system's center, as illustrated in Fig.~\ref{fig:7}(d). Since the external magnet lies on the central symmetry axis, the angular distribution of \(\Delta B_r\) (upper subplot) exhibits half-wave symmetry. Peak values occur at \(\theta = 0\) (sensor No.3) and \(\uppi\) (sensor No.6), in agreement with simulation results.  
 As observed in previous experiments, residual yoke magnetization affects the measurements: as the external magnet recedes, \(\Delta B_r\) stabilizes at non-zero values. The maximum \(\Delta B_r\) of \(3.5\,\mathrm{\upmu T}\) occurs at \(z = 155\,\mathrm{mm}\) and \(\theta = \uppi\), yielding \(\gamma_{\mathrm{max}} = 1.3 \times 10^{-5}\).  
 The lower subplot presents \(\Delta \overline{B}_r\). Notably, \(\Delta \overline{B}_r\) remains \(<\pm0.5\,\mathrm{\upmu T}\) regardless of the external magnet's position, confirming the simulation predictions.

\subsection{Discussion}

\begin{figure}[tp]
    \centering
    \includegraphics[width=0.4\textwidth]{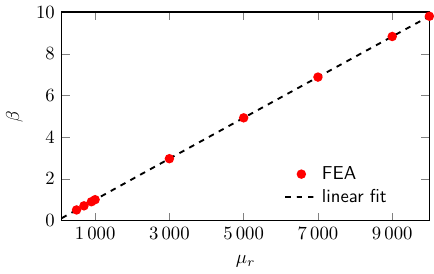}
    \caption{The correspondence between relative magnetic permeability and relative attenuation coefficient. The $\gamma(\mu_r = 1000)$ is selected as the reference.}
    \label{fig:8}
\end{figure}

 A comparison between FEA and experimental results reveals that the experimentally obtained attenuation factor is approximately one order of magnitude higher than the FEA-derived value. This discrepancy arises because the experimental magnetic flux within the yoke is significantly smaller, causing the yoke's magnetic permeability to approach its initial permeability. According to \cite{THUmag}, the initial relative permeability ($\mu_r$) is only approximately 1 000, whereas the simulation assumes $\mu_r = 10000$ close to the working permeability with the permanent magnets are integrated into the circuit. {In order to confirm the experimental result, here we also conducted FEA simulations with $\mu_r=1 000$. The results are presented in the second and third columns in Table~\ref{tab:experimental result}. It can be seen that} with the exception of the $\mathcal{H}^{r=175\,\mathrm{mm}}_{z=0\,\mathrm{mm}}$ measurement, which exhibits approximately twice the FEA-predicted value, all other experimental results show good agreement with computational predictions.
 {In actual operations of a Kibble balance, the magnetic circuit contains permanent magnets, and in this case the yoke permeability is higher than its initial permeability. With PMs, the magnetic shielding performance should be considerately improved compared to the experimental result. To characterize the shielding performance change on the yoke permeability, using the attenuation factor with $\mu_r = 1000$ as a reference point, the following conversion factor is defined,}
\begin{equation}
    \beta = \frac{\gamma(\mu_r = 1,000)}{\gamma(\mu_r)}.
    \label{eq:ur_conversion}
\end{equation}
Fig.~\ref{fig:8} illustrates how $\beta$ varies with increasing $\mu_r$ through FEA calculations. {It can be concluded that the attenuation factor, $\gamma$, is inversely proportional to the yoke permeability. Taking the Tsinghua system as an example, $\beta=9.98$, the attenuation factor $\gamma(\mu_r)$ is only one tenth of $\gamma(\mu_r=1000)$. The last column of Table~\ref{tab:experimental result} converts the experimental attenuation factor $\gamma(\mu_r)$ to the actual Kibble balance working condition, which is comparable to the FEA result with $\mu_r=10000$.} 

\begin{table}[]
    \centering
    \caption{The conversed experimental results of the attenuation rate}
    \renewcommand{\arraystretch}{1.8}
     \begin{tabular}{>{\centering\arraybackslash}p{0.09\textwidth} >{\centering\arraybackslash}p{0.1\textwidth} >{\centering\arraybackslash}p{0.1\textwidth} >{\centering\arraybackslash}p{0.1\textwidth}}
     \hline
     \hline
        Type & $\gamma_{\rm{max}}$, {FEA, $\mu_r=1000$} & $\gamma_{\rm{max}}$, {Exp. without PMs} & $\gamma_{\rm{max}}$, {KB working status} \\
     \hline
         $\mathcal{V}^{r=175\,\mathrm{mm}}_{z=-75\,\mathrm{mm}}$ & {$7.4 \times 10^{-5}$} & $5.9 \times 10^{-5}$ & $6.0 \times 10^{-6}$\\
         $\mathcal{H}^{r=175\,\mathrm{mm}}_{z=0\,\mathrm{mm}}$ & {$7.7 \times 10^{-5}$} & $1.4 \times 10^{-4}$ & $1.4 \times 10^{-5}$\\
         $\mathcal{V}^{r=0\,\mathrm{mm}}_{z=155\,\mathrm{mm}}$ & {$3.9 \times 10^{-5}$} & $2.0 \times 10^{-5}$ & $2.0 \times 10^{-6}$\\
         $\mathcal{H}^{r=0\,\mathrm{mm}}_{z=155\,\mathrm{mm}}$ & {$2.4 \times 10^{-5}$} & $1.3 \times 10^{-5}$ & $1.4 \times 10^{-6}$\\
     \hline
     \hline
    \end{tabular}
    \label{tab:experimental result}
\end{table}

\section{External magnetic flux error evaluation in the Tsinghua system}
\label{sec06}

 In this section, we take the Tsinghua tabletop Kibble balance \cite{THUmag} as an example and evaluate how much error could be introduced if the magnet-moving measurement scheme is used. Two major external magnetic flux sources are considered: 1) the geomagnetic field as a far-end flux source, and 2) the magnetic field created by the magnet-moving motor during velocity measurement, which is a near-end flux source. 

 {\subsection{External magnetic field measurement} \label{sec6-1}}
 
  As mentioned in section \ref{sec02}, in order to accurately measure external magnetic flux, it is necessary to exclude the leakage magnetic field of the magnet system. Here, we eliminated interference from the magnet system's leakage field by employing a modified assembly without permanent magnets. This approach preserves the system's response to external fields while effectively removing the leakage field contribution.

\begin{figure}[t]
    \centering
    \includegraphics[width=0.45\textwidth]{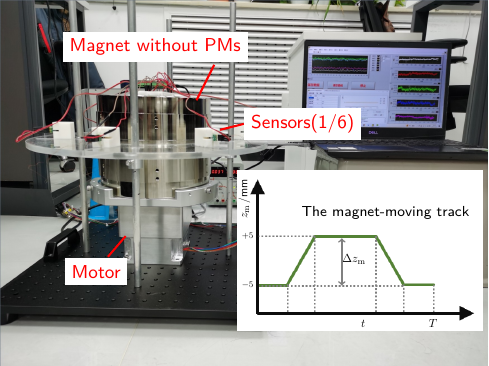}
    \caption{The experimental setup for the evaluation of the external magnetic environment. The subplot shows the moving track of the magnet system along the $z$ axis.}
    \label{fig:9}
\end{figure}

\begin{figure*}[tp!]
    \centering
    \includegraphics[width=0.88\textwidth]{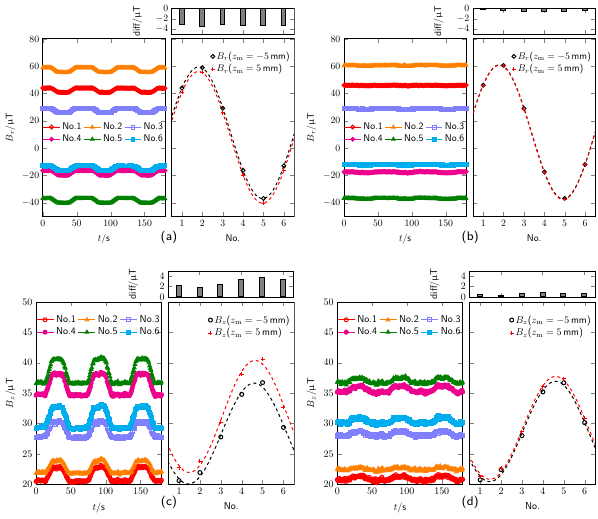}
    \caption{Measurement results of the external environmental magnetic field. (a) and (b) are the results of the radial magnetic flux density with moving only the magnet and moving both the magnet and sensors, respectively. (c) and (d) are the measurement results of the vertical field component with moving only the magnet and moving both the magnet and sensors, respectively. In each figure, the left subplot presents the measurements of the external $B_r$ over three cycles of magnet motion; the right lower subplot shows a sinusoidal fit of the radial magnetic field of $B_r(z_{\mathrm{m}}=-5\,\mathrm{mm})$ and $B_r(z_{\mathrm{m}}=5\,\mathrm{mm})$; the right upper subplot gives the $B_r$ difference when the magnet system is at two positions.  
    }
    \label{fig:10}
\end{figure*}

 The experimental setup is as shown in Fig. \ref{fig:9}. The magnet-lifting motor and the magnet system without permanent magnets are installed in the same way as in the actual Kibble balance measurement. All the additional supporting structures are made of non-magnetic materials. In the measurement, the motor lifts the magnet along the vertical direction by 10\,mm (the magnet system's center position $z_{\mathrm{m}}$ changes from -5\,mm to 5\,mm) with a measurement cycle of approximately 60\,s. We monitored the radial magnetic field on the symmetry surface of the magnet system ($z=0$\,mm) and the vertical magnetic field near the bottom of the magnet system ($z=-85$\,mm). The distance between the magnet system and the flux-gate sensors is 50\,mm ($r=160$\,mm). The results are shown in Fig. \ref{fig:10}.

 Fig.~\ref{fig:10}(a) presents the experimental results of radial magnetic flux density ($B_r$) measured at $(r=160\,\mathrm{mm}, z=0\,\mathrm{mm})$ using six flux-gate sensors. The left subplot demonstrates the variation in radial magnetic flux with moving the magnet meanwhile keeping the sensors static. The magnetic field of different sensors exhibits significant changes ($\approx-3\,\upmu$T) during magnet elevation, as shown in the left subplot. Notably, these variations persist even after motor cessation, indicating they primarily result from magnet position changes rather than motor operation. The right subplot displays the angular distribution of the radial magnetic field surrounding the magnet system. Near the symmetry plane, the field distribution follows a sinusoidal pattern, characteristic of the magnet system's influence.
 To isolate the effect of magnet position changes, we conducted subsequent experiments with sensors moving synchronously with the magnet, as shown in Fig.~\ref{fig:10}(b). Maintaining a constant relative distance between the magnet and sensors reduced magnetic field variations to well below 1\,$\upmu$T. No considerable field fluctuations were observed during motor operation, demonstrating that the motor's contribution to horizontal magnetic field leakage is negligible and complies with the external magnetic flux specifications outlined in Section~\ref{sec04}.

 The lifting motor is positioned directly beneath the magnet system. As established in Section~\ref{sec04}, the $\mathcal{V}^{r=0\,\mathrm{mm}}_{z}$ causes a residual field $\Delta \overline{B}_r$. However, direct measurement of the magnetic field strength at the motor surface is precluded by its metal enclosure. In this experiment, the vertical magnetic field leakage from the motor can be inferred from $B_z$ measurements. Fig.~\ref{fig:10}(c) presents the vertical magnetic flux density ($B_z$) measurements obtained at $(r=160\,\mathrm{mm}, z=-85\,\mathrm{mm})$ using the six flux-gate sensors. The left subplot reveals a substantial variation in $B_z$ of averaging 2.8 $\upmu$T when the magnet moves relatively to the sensors changes from $z_{\mathrm{m}} = -5\,\mathrm{mm}$ to $z_{\mathrm{m}} = 5\,\mathrm{mm}$. The right subplot displays the angular distribution of $B_z$, which unexpectedly approximates a sinusoidal function with a bias. Complementary measurements with moving sensors are shown in Fig.~\ref{fig:10}(d). As anticipated, the $B_z$ variation is markedly reduced, with an average difference of just 0.6\,$\upmu$T.

 The measurement detected no significant step change in the magnetic flux density during motor state transitions. This observation suggests that the $B_z$ component attributable to motor magnetic flux leakage is smaller than the spatial magnetic field gradient at the measurement location. We therefore employ the results from Fig.~\ref{fig:10}(d) to establish an upper bound for the motor's leakage field strength. The measured average $B_z$ variation during motor operation was 0.64 $\upmu$T, implying the motor-induced leakage field ($B_{z-\mathrm{mo}}$) must satisfy $B_{z-\mathrm{mo}} < 0.64\,\upmu$T. Reference to Table~\ref{tab:1} shows this value is substantially smaller than the established upper limit of $\mathcal{V}^{r=160\,\mathrm{mm}}_{z=-90\,\mathrm{mm}}$ (0.65 mT). Consequently, in the Tsinghua Kibble balance configuration, the motor's contribution to external magnetic flux introduces negligible measurement bias.

 \begin{figure*}[tp!]
     \centering
     \includegraphics[width=0.95\textwidth]{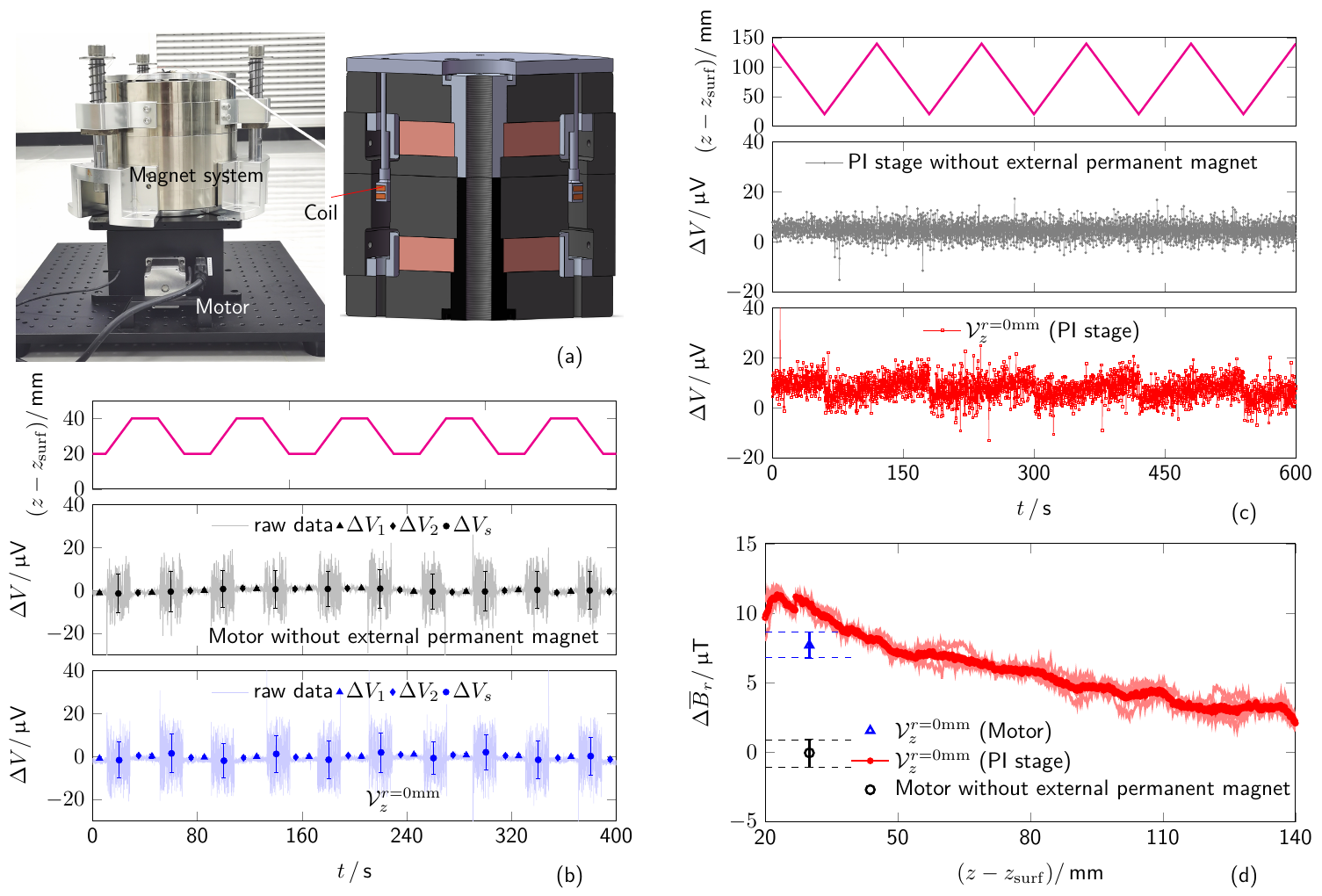}
     \caption{
     Measurement of voltage induced by external magnetic flux. (a) shows the experimental setup and coil suspension. (b) presents the measurement result with moving the motor and coil together. Subplots from top to bottom: movement trajectory of the motor, induced voltage without an external permanent magnet and induced voltage with the external magnet $\mathcal{V}{z}^{r=0\,\mathrm{mm}}$. 
     (c) presents the measurement result using a PI linear stage. From top to bottom: movement trajectory of the external permanent magnet, background signal without the magnet and induced voltage with the magnet $\mathcal{V}{z}^{r=0\,\mathrm{mm}}$. (d) calculates $\Delta\overline{B}r$ from the induced voltage measurements. The black circle represents results from motor operation without an external magnet. The blue triangle shows results from motor operation with $\mathcal{V}{z}^{r=0\,\mathrm{mm}}$. The red curve displays results from PI stage movement with $\mathcal{V}_{z}^{r=0\,\mathrm{mm}}$. The light red curve is the measurement in a single movement cycle (5 curves in total) and the dark red is their average.
     }
     \label{fig:11}
 \end{figure*}
 
 \subsection{Induced voltage test}

{A subsequent experiment replicated the Kibble balance's operational conditions using its intact magnet system as shown in Fig. \ref{fig:11}(a). In this setup, the coil was rigidly mounted to the magnet's top cover, forcing both to move concurrently when actuated by the motor. This configuration eliminates relative displacement between the coil and the main magnet, theoretically nullifying any induced voltage from the internal magnetic field. Therefore, the measured voltage $\Delta V$ is attributed solely to external magnetic flux sources, such as the geomagnetic field and motor stray fields. The experimental coil consisted of 350 turns of copper wire with an average radius of 85\,mm. The motor drove the assembly at a constant velocity of $v=1$\,mm/s over a 20\,mm range. The induced voltage was measured directly using a Keysight 3458A.
	
	The upper subplot of Fig.~\ref{fig:11}(b) depicts the trajectory of the motor's lifting movement. Five full measurement cycles, each lasting 40\,s, were performed. A 10-s static measurement of the coil voltage was recorded before and after each upward or downward movement. The results of these measurements are presented in the second row of Fig.~\ref{fig:11}(b). During movement, the signal-to-noise ratio of the induced voltage $\Delta V$ decreases significantly, mainly due to coil variations. The average induced voltage during movement is denoted as $\Delta V_s$, while the averages from the static periods before and after movement are $\Delta V_1$ and $\Delta V_2$, respectively. The net induced voltage is calculated as
	\begin{equation}
		\Delta\overline{V}(i)=\Delta V_s(i)-\frac{\Delta V_1(i)+\Delta V_2(i)}{2},
		\label{delV}
	\end{equation}
	where $i=1,2,...,10$ denotes the sequence number for each movement. This step removes the majority of the voltmeter offset drift that occurs during the 40-s cycle.
	A subsequent subtraction of adjacent sequences gives the induced voltage difference of opposite movement directions ($\pm v$), and hence the induction in a single movement is written as
	\begin{equation}
		\delta V(j)=\frac{\Delta\overline{V}(2j)-\Delta\overline{V}(2j-1)}{2},
	\end{equation}
	where $j=1,2,...,5$. $\delta V$ represents the induced voltage from external flux, and it can be converted into a magnetic field change at the coil position using $\Delta \overline{B}_r = \delta V/(l v)$, where the coil wire length is $l \approx 186.9$\,m. 
	
	Analysis of the results (Fig.~\ref{fig:11}(b), middle row) yields a five-point average, $\delta \overline{V} = (0.006 \pm 0.188)\,\upmu$V. This corresponds to a magnetic field change of $\Delta \overline{B}_r = (0.03 \pm 1.01)\,\upmu$T. The measured signal is theoretically negligible based on the surrounding magnetic field measurements in Section~\ref{sec6-1}. Although the $\Delta \overline{B}_r$ is on the order of $10^{-8}$\,T, its uncertainty ($\approx1\,\upmu$T) is too large to confirm that external flux from the motor has a negligible effect on the measurement.

	Although direct measurement of how close $\Delta \overline{B}_r$ is to zero is limited by voltage measurement noise, we can circumvent this limitation by conducting experiments with an amplified external flux source. Based on FEA simulations, the configuration $\mathcal{V}_{z}^{r=0\,\mathrm{mm}}$ has the most significant impact on shielding performance. Consequently, an external field setup analogous to those in the left column of Fig. \ref{fig:5} and Fig. \ref{fig:7}(c) was applied. Note that, limited by the motor movement range, $z-z_\mathrm{surf}$ was set from 20\,mm to 40\,mm. The measurement results are shown in the lower row of Fig. \ref{fig:11}(b). In this amplified setup, a clear induced voltage signal was detected, yielding $\delta \overline{V}=(1.439\pm0.182)\,\upmu$V and $\Delta \overline{B}_r=(7.70\pm0.97)\,\upmu$T. This result agrees with the FEA calculation in Fig. \ref{fig:5}(d) (the average value over the range from 20\,mm to 40\,mm).
	As shown in Section \ref{sec6-1}, the motor itself introduces an external flux below $1\,\upmu$T during movement. Given that the surface field strength of the permanent magnet used for amplification is approximately 200\,mT, the scaling factor between the two scenarios is approximately $5 \times 10^{-6}$. By applying this factor, the field change $\Delta \overline{B}_r$ induced by the motor alone (without the external magnet) is estimated to be well below $1 \times 10^{-9}$.
	
	To mitigate interference from coil vibration caused by motor operation, we implemented an alternative experimental configuration: the magnet system and coil were kept stationary while an external permanent magnet ($\mathcal{V}^{r=0\,\mathrm{mm}}_{z}$) was moved directly above them. The magnet was translated with uniform motion using a PI stage (PI M-413.2DG) at a velocity of 2\,mm/s over a range from 20\,mm to 140\,mm above the upper surface of the magnet system. The moving trajectory of the external permanent magnet is illustrated in the upper subplot of Fig. \ref{fig:11}(c).
	The middle subplot displays the induced voltage measured during PI stage operation without the external permanent magnet, which serves as the background signal for subsequent data processing. The lower subplot shows the induced voltage measured with the external magnet moving. After subtracting the background signal (middle subplot) and performing data analysis to extract $\delta \overline{V}$ and $\Delta \overline{B}_r$, the $\Delta \overline{B}_r(z)$ curve was represented by the red curve in Fig. \ref{fig:11}(d).
	The measured $\Delta \overline{B}_r(z)$ result is consistent with the FEA calculations shown in Fig. \ref{fig:5}(d). Specifically, at $(z-z_{\mathrm{surf}}) = 20$\,mm, $\Delta\overline{B}_r$ is approximately 10\,$\upmu$T. For comparison, measurement results from moving the magnet and the coil are also included in Fig. \ref{fig:11}(d). The results of two external permanent magnet experiments are comparable, which further confirms the validity of the induced voltage test.

	In summary, the experiments presented in this section demonstrate that actively monitoring the external magnetic flux density — keeping it within the established field limits — suppresses the related measurement error in the Kibble balance to below $1 \times 10^{-8}$. This strategy of forgoing additional magnetic shielding facilitates the construction of more compact Kibble balances and simplifies their operation.}

\section{Conclusion}
\label{sec07}
 For magnet-moving Kibble balances, addressing measurement errors induced by external magnetic flux is crucial. Direct evaluation of these errors proves challenging due to interference from the main magnetic field generated by permanent magnets. This paper presents an effective evaluation methodology based on the attenuation factor $\gamma$, supported by FEA.
 The proposed approach systematically examines the coupling pathways, distribution pattern, and influence mechanism of external magnetic flux into the magnet system, as well as a quantitative determination of attenuation factors and external flux limitations. Key findings of particular significance include that vertical far-end magnetic flux affects the weighing phase measurement. Since this error correlates linearly with coil position, it can be effectively mitigated through strategic positioning adjustments. Near-end magnetic flux exhibits distinct critical regions: Vertical components along the magnet's vertical symmetrical axis $z$ induce a maximal error. Horizontal components at the symmetry plane ($z = 0$\,mm) produce peak disturbance. These regions should be protected from external magnetic interference.
 
 We assembled a prototype magnet system without permanent magnets to experimentally validate the proposed methodology. The experimental results demonstrate an attenuation factor on the order of $10^{-6}$ for the near-end external field and a negligible magnetic flux leakage from the motor in the Tsinghua Kibble balance system, well within environmental constraints. {Experimental tests were also conducted based on induced voltage measurements from a coil fixed to the magnetic circuit. In this configuration, the permanent magnets are integrated into the system, closely approximating the actual working conditions of a Kibble balance. The results further confirm the FEA calculations and agree with the fluxgate magnetometer measurements. A direct comparison of the $Bl$ value using two different magnet-coil movement mechanisms is planned for future work once the Kibble balance is fully operational.}

\end{document}